\documentclass[journal]{IEEEtran}

\usepackage{tablefootnote}
\usepackage{enumitem,kantlipsum}
\usepackage{graphicx}
\usepackage{subcaption}
\usepackage{tikz}
\usepackage{comment}

\usepackage{tabularx}
\usepackage{arydshln}

\usepackage{amsmath}

\usepackage{adjustbox}

\usepackage{cite}

\usepackage{multicol}
\usepackage{multirow}

\usepackage{url}
\usepackage{hyperref}

\begin{document}
%
\title{Architecture Analysis and Benchmarking of 3D U-shaped Deep Learning Models for Thoracic Anatomical Segmentation}

\author{Arash Harirpoush,
        Amirhossein Rasoulian,
        Marta Kersten-Oertel,
        and Yiming Xiao,~\IEEEmembership{Senior Member,~IEEE}
\thanks{A. Harirpoush, A. Rasoulian, M. Kersten-Oertel, and Y. Xiao are with the Department of Computer Science and Software Engineering, Concordia University, Montreal, Canada. Email: aharirpoosh@gmail.com}
}

\maketitle

\begin{abstract}
Recent rising interests in patient-specific thoracic surgical planning and simulation require efficient and robust creation of digital anatomical models from automatic medical image segmentation algorithms. Deep learning (DL) is now state-of-the-art in various radiological tasks, and U-shaped DL models have particularly excelled in medical image segmentation since the inception of the 2D UNet. To date, many variants of U-shaped models have been proposed by the integration of different attention mechanisms and network configurations. Systematic benchmark studies which analyze the architecture of these models by leveraging the recent development of the multi-label databases,  can provide valuable insights for clinical deployment and future model designs, but such studies are still rare. We conduct the first systematic benchmark study for variants of 3D U-shaped models (3DUNet, STUNet, AttentionUNet, SwinUNETR, FocalSegNet, and a novel 3D SwinUnet with four variants) with a focus on CT-based anatomical segmentation for thoracic surgery. Our study systematically examines the impact of different attention mechanisms, the number of resolution stages, and network configurations on segmentation accuracy and computational complexity. To allow cross-reference with other recent benchmarking studies, we also included a performance assessment of the BTCV abdominal structural segmentation. With the STUNet ranking at the top, our study demonstrated the value of CNN-based U-shaped models for the investigated tasks and the benefit of residual blocks in network configuration designs to boost segmentation performance.

\end{abstract}


\IEEEpeerreviewmaketitle

\section{Introduction}
\IEEEPARstart{I}{n} modern surgical planning that emphasizes high precision and low trauma, 3D anatomical segmentation from pre-operative medical images is becoming increasingly important. Thoracic surgery, i.e., chest surgery which involves operations on lungs affected by cancer, trauma, pulmonary disease, or cardiac conditions, accounts for approximately 530,000 cases per year in the US \cite{byrd2022brief}. In addition to video-based surgical guidance with limited spatial information, recent studies \cite{qiu2020three, ghosh2022clinical} have demonstrated significant advantages of using patient-specific physical or digital 3D models for various thoracic surgeries \cite{qiu2020three, ghosh2022clinical} in both conventional and mixed reality surgical environments \cite{ujiie2021developing, chan2015preoperative, sardari2019interactive}. To ensure the outcomes of these applications, efficient and accurate 3D anatomical segmentation and reconstruction is essential. 

Deep learning (DL) approaches, such as convolutional neural networks (CNNs) have dominated the state-of-the-art performance in various radiological tasks. With their quick inference time, they offer a tool to enable efficient digital twin construction for thoracic surgical planning, simulation, and intra-operative monitoring. U-shaped models, pioneered by the 2D UNet \cite{ronneberger2015u}, stand out among DL segmentation models \cite{liu2021review} for their robust performance and elegant architecture. The typical U-shaped architecture comprises three key elements: an encoder for learning relevant image features and compressing them into lower-dimensional embeddings, a decoder for expanding these embeddings and producing the final segmentation, and skip connections that maintain fine-grained details during upsampling by aggregating feature maps across encoder and decoder layers. Since the first inception, major efforts have been dedicated to adapting the 2D framework to 3D \cite{cciccek20163d}, exploring different backbones for the encoder/decoder \cite{isensee2021nnu, huang2023stu, oktay2018attention, hatamizadeh2022swin, rasoulian2023weakly,cao2022swin}, updating the resolution stages \cite{yousef2023u}, and experimenting with novel network configurations \cite{gut2022benchmarking}. To better understand the impact of these enhancements and investigate the application of 3D anatomical reconstruction for thoracic surgical planning and simulation, a comprehensive benchmark study, providing the models' architectural characteristics, would be highly instrumental, but has yet to be conducted.

\begin{figure*}
    \centering
    \includegraphics[width=\textwidth]
    {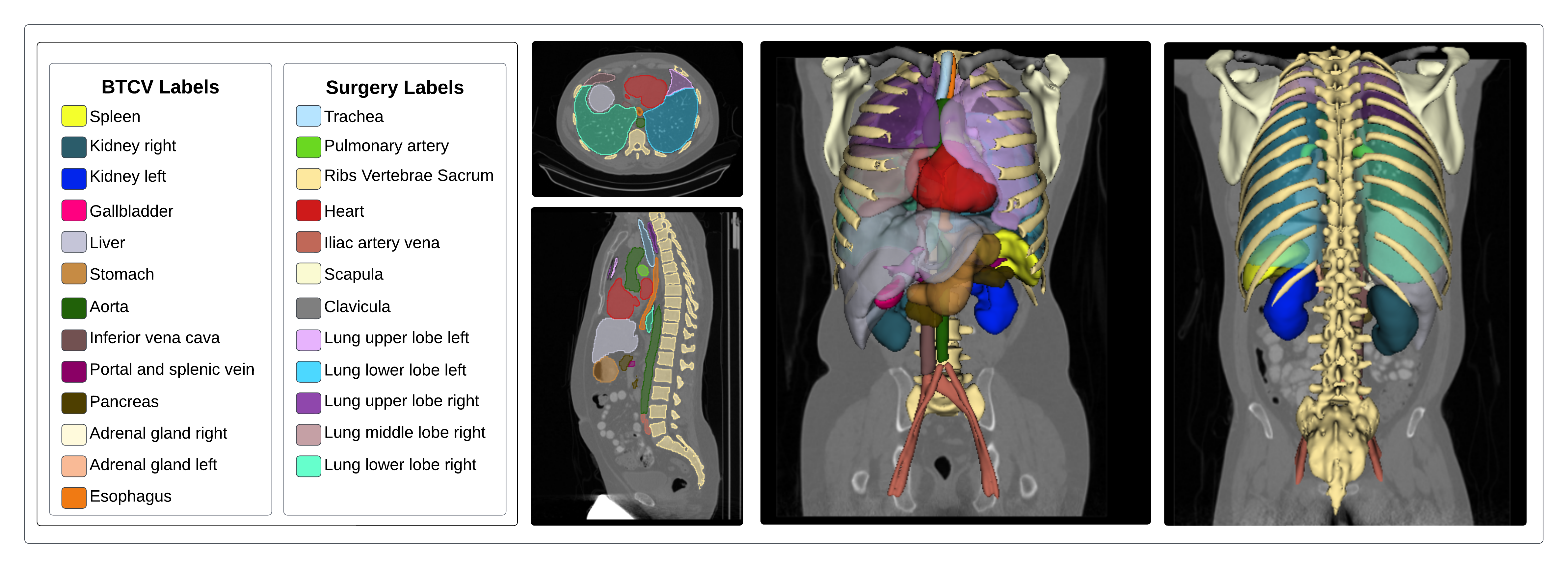}
    \caption{Demonstration of the anatomical structures for U-shaped model benchmarking, including 12 labels for thoracic surgery and 13 labels that are consistent with the BTCV segmentation challenge.}
    \label{fig:Labels}
\end{figure*}

\subsection{Variants of U-shaped Architectures}
Many variants of the U-shaped architectures have been proposed primarily for medical image segmentation \cite{ronneberger2015u}. Among these, the nnUNet\cite{isensee2021nnu} framework, which allows task-specific optimization of 2D and 3D UNet models and training strategies, has achieved great success in a wide range of segmentation tasks. Recently, Huang \emph{et al.}~\cite{huang2023stu} proposed the STUNet, an enhancement of 3DUNet  model from the nnUNet\cite{isensee2021nnu} framework, with modified downsampling and upsampling blocks in the encoder and decoder, respectively. 

Besides architecture and training strategy optimization, some attempts have also been made to incorporate various attention mechanisms in U-shaped models. In the AttentionUNet\cite{oktay2018attention}, attention gates that combine trained soft attention feature maps through skip connections are implemented to enhance accuracy and model transparency for pancreas segmentation. With the emergence of the Vision Transformer (ViT), which leverages self-attention to capture long-range dependencies within an image, CNN-Transformer-hybrid U-shaped models were introduced to enhance the performance of the CNN-based UNet. 

For example, the TransUNet\cite{chen2021transattunet}, CoTR\cite{xie2021cotr}, and TranBTS\cite{wang2021transbts} models strategically incorporated Transformer layers into the concluding stages of the encoder to enable enhanced features extraction from the feature maps generated by the preceding CNN blocks. More recently, the UNETR\cite{hatamizadeh2022unetr} and SwinUNETR \cite{hatamizadeh2022swin} advocate for utilizing fully-transformer-based encoders by using ViT and Swin Tranformer \cite{liu2021swin}, respectively, while retaining CNN decoders. 
Attempts have also been made to create U-shaped models driven fully by Transfomer blocks. Notably, the 2D Swin-UNet\cite{cao2022swin} first implemented this approach and demonstrated its performance on abdominal CT segmentation. Similarly, the VT-UNet\cite{peiris2022robust} adopts the same approach, with further addition of cross-attention mechanisms in the decoder. In the same category, Zhou \textit{et al.} \cite{zhou2023nnformer} proposed the nnFormer, which introduced local and global self-attention mechanisms in the encoder, decoder, and skip-attentions. Lastly, the more recent Focal Modulation\cite{yang2022focal} offers an alternative attention mechanism that models hierarchical contextualization of image features, which are aggregated for each query token. This relatively new technique was shown to outperform its state-of-the-art counterparts, such as Swin and focal Transformers in 2D natural image recognition and segmentation tasks. Leveraging this new mechanism, the FocalSegNet \cite{rasoulian2023weakly} replaces the Transformer blocks of the encoder in the UNETR with new 3D Focal Modulation blocks to perform volumetric medical image segmentation. 

Aside from attention mechanisms in U-shaped models, earlier research has explored various setups for the number of resolution stages and diverse skip connection schemes \cite{zhou2018unet++, xiang2022towards, wang2022uctransnet, chen2021transattunet} to improve segmentation accuracy. Among these, UNet++\cite{zhou2018unet++} introduced nested hierarchical skip connections to fuse encoder and decoder features, and the BiO-Net \cite{xiang2022towards} and its variants leveraged bidirectional skip connections within the U-Net model. Another technique in this context is to incorporate Transformers in skip connections for 2D medical image segmentation that is more computationally intensive \cite{wang2022uctransnet, chen2021transattunet}.

\subsection{Previous Anatomical Segmentation Benchmarking studies}
The increasing amount of public datasets, such as Kits\cite{heller2019kits19}, CHAOS\cite{kavur2021chaos}, SegTHOR\cite{lambert2020segthor}, BTCV\cite{landman2015miccai}, and ACDC\cite{bernard2018deep} have greatly facilitated the benchmarking of various new algorithms in medical image segmentation. For 2D OCT image segmentation, Kugelman \textit{et al.} \cite{kugelman2022comparison} tested eight U-shaped models (VanillaUNet, DenseUNet, AttentionUNet, SEUNet, ResidualUNet, R2UNet, UNet++, and InceptionUNet), and concluded that the VanillaUNet was the most appropriate considering computational complexity while all models performed similarly. Yao \textit{et al.} \cite{yao2023cnn} benchmarked UNet, UNet++, TransUNet, and SwinUnet for 2D segmentation tasks of the lung and abdominopelvic ovarian masses, and they showed that the TransUNet model had the best performance in both datasets. In their paper, Ji \emph{et al.} introduce the AMOS\cite{ji2022amos} MRI and CT dataset for 3D abdominal anatomy segmentation and tested UNet, VNet\cite{milletari2016v}, CoTr,  UNETR, SwinUNETR, and nnFormer on the proposed dataset. When considering both accuracy and model complexity (i.e., model parameters and the number of GFlops), they found no advantage of using transformers over CNNs, with the UNet achieving the best accuracy, particularly for larger anatomies. After reviewing various U-shaped transformer-based models in 2D and 3D medical image segmentation, Xiao \textit{et al.} \cite{xiao2023transformers} evaluated Swin-Unet, LeViT-UNet, UCTransNet, TransAttUnet, UTNet, and UTNetV2 on MSD \cite{antonelli2022medical} and NSCLC \cite{bakr2018radiogenomic} datasets for 2D segmentation of pancreas and lung cancer, respectively. Their study showed that while UTNetV2 is the best for pancreas segmentation, TransUNet achieved the best performance in lung cancer segmentation. They also found that combining CNN and Transformer was beneficial, and the choice of the best model can be task-specific.

\subsection{Novelty and Contributions}

While several groups have benchmarked U-shaped models as systematic studies or validation of a new algorithm, there are still remaining research gaps that we aim to address in this study. First, few studies investigated 3D segmentation of the lung \cite{xiao2023transformers} and other related anatomies for thoracic surgical planning and simulation. Second, many studies focused on 2D segmentation tasks and models while direct 3D processing can offer better spatial continuity across slices in the segmented labels. Finally, there is a lack of dedicated exploration for the impact of various attention mechanisms and network configuration designs on multi-organ 3D segmentation tasks. To tackle these, our study evaluates the performance of several U-shaped models, including 3DUNet, STUNet, 3D AttentionUNet, SwinUNETR, FocalSegNet, and a new 3DSwinUnet along with its variants, in segmenting anatomical structures associated with thoracic surgery from CT scans of the TotalSegmentator dataset\cite{wasserthal2022totalsegmentator}. To allow an easy comparison of conclusions with concurrent literature, we also benchmarked these models for the anatomies in the BTCV\cite{landman2015miccai} challenge in the same dataset (see Fig. \ref{fig:Labels}). Our main contributions can be summarized as follows:

\begin{itemize}[leftmargin=*]

    \item Investigate the effects of different attention mechanisms (attention-gate, self-attention, focal modulation, and baseline UNet) in U-shaped architectures

    \item Benchmark state-of-the-art U-shaped models, including our novel 3D SwinUNet and its four variants in 3D image segmentation for thoracic surgery and BTCV challenge anatomies by considering their accuracy and computational complexity
    
    \item Investigate the impacts of different network configuration designs, including the number of resolution stages (number of upsampling/downsampling operations) and different operations for skip connections, downsampling, and upsampling in a pure Swin Transformer-based U-shaped model
    
    \item Provide the code and trained model weights from the study in an open-access repository (\href{https://github.com/HealthX-Lab/DeepSegThoracic}{https://github.com/HealthX-Lab/DeepSegThoracic}) to the research and clinical communities
\end{itemize}

\section{Methods and Materials}

\subsection{Dataset}\label{dataset}
We employed the dataset from the TotalSegmentator paper\cite{wasserthal2022totalsegmentator}, which included 1204 body CT scans with 104 different labels. These labels covered 27 organs, 59 bones, 10 muscles, and 8 vessels. For our study, we focused on 79 annotations that are relevant to thoracic surgery and correspond to the BTCV challenge. As there is no need for further division of some anatomies and to facilitate training and evaluation, we combined some of the annotations that belong to the same anatomical structure, resulting in 25 labels: 12 for thoracic surgery and 13 that correspond to the BTCV challenge. The list and visualization of these anatomies are shown in Fig. \ref{fig:Labels}. As the CT scans in the dataset contain different fields-of-view and an inconsistent number of labels, we selected the cases that contain more than 22 of the 25 classes that we defined, resulting in 440 training cases, 26 validation cases, and 30 testing cases. To facilitate the assessment, we also provide the distributions of each anatomical label across the training and testing datasets in Fig. S1 of the \textit{Supplementary Materials}.

\subsection{Experimental set-up and network training}
We selected six U-shaped models for performing the benchmarking, including a 3DUNet optimized based on the nnUNet framework\cite{isensee2021nnu}, STUNet\cite{huang2023stu}, AttentionUNet\cite{oktay2018attention}, Swin-UNETR\cite{hatamizadeh2022swin}, FocalSegNet\cite{rasoulian2023weakly}, and a new 3D implementation of the original 2D Swin-Unet\cite{cao2022swin} (we refer to as 3DSwinUnet). A visual summary of these models' architectures is included in Fig. S2 of the \textit{Supplementary Materials}. Furthermore, we also devised 4 additional variants of the 3DSwinUnet with different strategies for skip connections, feature map downsampling, and upsampling (more details in Section \ref{methods: ablation}). Such a range of models allows us to comprehensively assess the impact of different attention mechanisms, as well as more nuanced network configuration designs for U-shaped models. Specifically, the STUNet \cite{huang2023stu} and FocalSegNet\cite{rasoulian2023weakly} models were taken from their github repositories while implementation of 3D AttentionUNet and SwinUNETR models in MONAI \cite{cardoso2022monai} were used. Finally, the 3DSwinUnet and its variants were implemented using the MONAI framework. The 3DUNet, STUNet, and AttentionUNet models contain five resolution stages while the SwinUNETR and FocalSegNet models have four resolution stages. Finally, the 3DSwinUnet model and its variants consist of three resolution stages \cite{cao2022swin}. 

As the dimension of the input image patch is $96\times96\times96$ voxels across all selected neural network architectures, to allow a similar field-of-view per 3D input patch to the original TotalSegmentator paper \cite{wasserthal2022totalsegmentator} for comparison, we resampled the CT scans to a $2.0 \times 2.0 \times 2.0 mm^{3}$ resolution. For all model training, we used the SGD optimizer with an initial learning rate of 0.01, a Nesterov momentum of 0.99, and a weight decay of $1e-3$ to minimize the loss function which is a sum of the cross-entropy and Dice loss. For the learning rate scheduler, we employed the Poly method, which reduces the learning rate by increasing the number of epochs by calculating $LR \times (1 - Epoch/MaxEpoch)^{0.9}$, where $LR$ represents the initial learning rate, $Epoch$ is the current epoch number, and $MaxEpoch$ is the maximum epoch number. Each epoch consists of 250 iterations and a batch size of two. Finally, data augmentation techniques, including random rotation and scaling were added to enhance the robustness of training.

\subsection{Ablation studies} \label{methods: ablation}
In addition to comparing the six established U-shaped models, we also conducted ablation studies on them to further probe the relevant design choices that can influence their performance. First, to confirm the impact of attention mechanisms, we evaluated the variants with these DL models, all with four resolution stages. Second, to understand the impact of the number of resolution stages, we compared these models and their variants that are one stage different from the original architectures. Note that, here we evaluated 3DSwinUnet with three and four resolution stages using a patch embedding size of two, as the four-staged version has a downsampling limitation on the input patch size at the last resolution stage. Finally, the more nuanced design elements for U-shaped models, including the operations for skip-connections (SC), downsampling (DS), and upsampling (US) are often overlooked. To further improve the adapted 3DSwinUnet, which is a full Swin Transformer model and to better understand the effects of these operations, we investigated the influences of these design choices on the baseline 3DSwinUnet. Following the original 2D design, the baseline 3DSwinUnet employs linear layers in its skip-connection, downsampling, and upsampling. Here, we introduced four additional 3DSwinUnet variants, named 3DSwinUnetV1 $\sim$ 3DSwinUnetV4, with their detailed design differences listed in Table \ref{table:SwinUnetVarients}. For skip-connection and downsampling, we compared the application of linear and residual operations while for upsampling, we tested linear operation, nearest interpolation, and transpose convolution. 

\subsection{Evaluation Metrics and Statistical Analysis} \label{evaluation}
\noindent
For each CT scan of the test cases, sliding windows with 50\% overlapping were used to compute the automatic segmentation results, which were combined into one volume using a Gaussian weighting function with a standard deviation coefficient of 0.125. We compared the aforementioned DL models for their segmentation accuracy and efficiency for the Surgical Labels and BTCV Labels, which include anatomical structures of various sizes and geometries. Specifically, in terms of segmentation accuracy, we used the Dice coefficient and Normalized Surface Distance (NSD), both with the range of [0,1] (1 being the most desirable). While the first metric was widely used, it tends to favor larger objects and those with a bigger surface-to-volume ratio. To complement the Dice coefficient, the NSD measures the tightness of fit for the surfaces between the automatic segmentation and the ground truths. It can more appropriately assess the segmentation quality for smaller anatomies. Here, we used a $3 mm$ threshold for computing the NSD. In terms of computational complexity, we evaluated the number of parameters in the model and the inference latency. Specifically, the inference latency is the amount of time needed for a model to process one $96 \times 96 \times 96$ voxel image patch and was obtained by using the benchmark framework with 1000 run time provided by \cite{hedegaard2022pytorchbenchmark} on a desktop computer with an NVIDIA GeForce RTX 3090 GPU and an 11th Gen Intel® Core™ i9 CPU. 

To verify the differences in segmentation performances across different models, we performed statistical analysis on the results. Specifically, two-way ANOVA tests were conducted to investigate whether group-wise difference exists among the tested models, and whether there exists a difference in terms of segmentation performance between the Surgical labels and BTCV labels. If the ANOVA test indicates a significant difference, pair-wise post-hoc analysis with Tukey's Honestly Significant Difference (HSD) was employed to further identify the between-model differences with statistical significance. Here, we defined a p-value of 0.05 as the threshold that indicates a statistical significance.

\subsection{Algorithm Ranking Method}
To properly assess the performance of algorithms \cite{maier2018rankings}, public medical image processing challenges have widely adopted various ranking methods that incorporate multiple evaluation metrics. In general, two main ranking mechanisms have been utilized in algorithm ranking: 

\begin{itemize}[leftmargin=*]

\item \textit{Metric-based aggregation (``aggregate then rank")}: the evaluation metrics are first aggregated using median or mean across all cases, and then the ranking is based on the aggregated value. 
\item \textit{Case-based aggregation (``rank then aggregate"}: the models are first ranked for each metric on each case, and then the average values of all the rankings are used to rank the models. 
\end{itemize}

\noindent
Maier \textit{et al.} \cite{maier2018rankings} found that metric-based aggregation is the most commonly used ranking method in different medical image processing challenges, and recommend using it over case-based aggregation. Furthermore, for metric-based aggregation, the mean as an aggregation method was evaluated to be more robust than the median. Following these guidelines, we adopted the ``aggregate then rank" approach with the mean value for metric aggregation. Specifically, we first aggregated the results of all test cases and then ranked the models for each metric. With this strategy, we obtained the algorithm rankings for segmentation accuracy and model complexity by averaging the rankings of the sub-measures of each category. Finally, to fully consider both categories of factors, the final algorithm ranking was achieved based on the average of the segmentation accuracy ranking and model complexity ranking for each algorithm. For the assessment, we obtained the segmentation accuracy rankings and the final rankings for the Surgical label, BTCV labels, and the combination of both groups.

\section{RESULTS} \label{benchmark}

In Table \ref{table:classDice}, we summarize the segmentation accuracy, model complexity, and ranking for the 3DUNet, STUNet, AttentionUNet, SwinUNETR, FocalSegNet, and 3DSwinUnet. Additionally, the evaluation metrics of the ablation studies are listed in Tables \ref{table:FourStagePerfRank} to \ref{table:SwinUnetVarients}. Finally, due to the page limit, we demonstrate the segmentation outcomes of these models with image examples and anatomy-wise boxplots in Fig. S3-S4 and S5-S6 of the \textit{Supplementary Materials}, respectively.

\begin{table*}[htbp]
  \centering
  \renewcommand{\arraystretch}{2}
  \caption{Model's computational complexity, accuracy (mean$\pm$std), and ranking across U-shaped models. Note that an individual metric's rankings are shown as superscripts beside the corresponding metrics.}
  \label{table:classDice}
  \resizebox{\textwidth}{!}{
    \begin{tabular}{|c|c|c|c|c:c:c|c:c:c|c:c:c|c:c:c|}
      \hline
        \multirow{2}{*}{\textbf{Model Name}} & \multicolumn{3}{c|}{\textbf{Model Complexity $\downarrow$}} & \multicolumn{3}{c|}{\textbf{Dice $\uparrow$}} & \multicolumn{3}{c|}{\textbf{NSD $\uparrow$}} & \multicolumn{3}{c|}{\textbf{Segmentation Rankings $\downarrow$}} & \multicolumn{3}{c|}{\textbf{Final Rankings $\downarrow$}}\\
        \cline{2-16} & Number of Parameter (M) & Inference Latency ($ms \pm ms$) &
        Ranking & BTCV Labels & Surgery Labels & Total & BTCV Labels & Surgery Labels & Total & BTCV & Surgery & Total & BTCV & Surgery & Total \\
         \hline
         $3DUNet$ & $30.6 ^4$ & $6.158 ^1$ & 1 & $93.66 \pm 3.08 ^3$ & $96.96 \pm 1.11 ^2$ & $95.18 \pm 1.69 ^2$ & $97.50 \pm 2.89 ^2$ & $98.88 \pm 1.25 ^1$ & $98.13 \pm 1.65 ^2$ & 2 & 1 & 2 & 2 & 1 & 2 \\
        \hline
        $STUNet$ & $30.23 ^2$ & $7.298 ^3$ & 1 & $94.08 \pm 2.92 ^1$ & $97.04 \pm 1.20 ^1$ & $95.44 \pm 1.62 ^1$ & $97.57 \pm 2.54 ^1$ & $98.85 \pm 1.59 ^2$ & $98.16 \pm 1.54 ^1$ & 1 & 1 & 1 & 1 & 1 & 1 \\
        \hline
        $AttentionUNet$ & $30.59 ^3$ & $7.193 ^2$ & 1 & $93.78 \pm 3.11 ^2$ & $96.76 \pm 1.29 ^3$ & $95.14 \pm 1.74 ^3$ & $97.40 \pm 2.84 ^3$ & $98.77 \pm 1.60 ^3$ & $98.03 \pm 1.69 ^3$ & 2 & 2 & 3 & 2 & 2 & 3 \\
        \hline
        $SwinUNETR$ & $62.19 ^6$ & $18.585 ^5$ & 3 & $93.32 \pm 3.21 ^5$ & $96.54 \pm 1.45 ^5$ & $94.79 \pm 1.83 ^5$ & $97.10 \pm 2.86 ^5$ & $98.28 \pm 1.81 ^5$ & $97.64 \pm 1.78 ^5$ & 4 & 4 & 5 & 4 & 4 & 5 \\
        \hline
        $FocalSegNet$ & $69.65 ^7$ & $15.412 ^4$ & 3 & $93.47 \pm 2.98 ^4$ & $96.57 \pm 1.39 ^4$ & $94.89 \pm 1.67 ^4$ & $97.20 \pm 2.81 ^4$ & $98.43 \pm 1.58 ^4$ & $97.77 \pm 1.66 ^4$ & 3 & 3 & 4 & 3 & 3 & 4 \\
        \hline
        $3DSwinUnet$ & $7.98 ^1$ & $22.91 ^6$ & 2 & $46.15 \pm 7.02 ^7$ & $59.30 \pm 4.60 ^7$ & $52.20 \pm 5.06 ^7$ & $34.81 \pm 5.64 ^7$ & $38.52 \pm 4.15 ^7$ & $36.54 \pm 4.24 ^7$ & 6 & 6 & 7 & 4 & 4 & 5 \\
        \hline
        \hline
        $3DSwinUnetV4$ & $31.55 ^5$ & $29.025 ^7$ & 4 & $92.04 \pm 3.58 ^6$ & $95.72 \pm 1.58 ^6$ & $93.73 \pm 2.00 ^6$ & $96.14 \pm 3.47 ^6$ & $97.50 \pm 1.96 ^6$ & $96.76 \pm 2.06 ^6$ & 5 & 5 & 6 & 5 & 5 & 6 \\
        \hline
        \end{tabular}%
  }
\end{table*}

\subsection{Segmentation Accuracy} \label{SegmentationAccuracy}
With the segmentation performance of the six U-shaped models, and the best-performing 3DSwinUnet (i.e., 3DSwinUnetV4) in Table \ref{table:classDice}, we found three general observations. First, these included models achieved significantly higher Dice scores on the surgical labels rather than BTCV ones (p$<$0.05). Second, the 3DSwinUnet model performed significantly worse in segmentation accuracy than the rest of the counterparts (p$<$0.05). Finally, despite the differences in model architectures, especially in the adoption of diverse attention mechanisms, and the variations in the mean metrics, their segmentation performances do not differ significantly (p$>$0.05). Specifically, for the surgical labels, the 3DUNet achieved the second best mean Dice score of 96.96\% and the best mean NSD of 98.88\% while the STUNet had the highest mean Dice score of 97.04\% and the second best NSD of 98.85\%, making them share the first ranking for segmentation accuracy in this task. The AttentionUNet, FocalSegNet, SwinUNETR, 3DSwinUnetV4, and 3DSwinUnet followed in our ranking, respectively. For the BTCV labels, the STUNet model also achieved the best Dice scores (94.08\%) and NSD (97.57\%). 3DUNet and AttentionUNet shared the second place in the ranking, with 3DUNet achieving the second best mean NSD (97.50\%) while AttentionUNet had the second best Dice score of (93.78\%). After them, the models in descending order of ranking are the FocalSegNet, SwinUNETR, 3DSwinUnetV4, and 3DSwinUnet. Finally, when pulling both sets of labels together, the STUNet model again achieved the highest Dice score of 95.44\% and the best NSD of 98.16\%, with the 3DUNet ranked second, followed by the AttentionUNet, FocalSegNet, SwinUNETR, 3DSwinUnetV4, and 3DSwinUnet. 

\subsection{Model Complexity}
Primarily due to the choice of architecture types, the model complexity varies, with the 3DUNet having the lowest inference latency. By sharing the basic structure of the UNETR\cite{hatamizadeh2022unetr}, the SwinUNETR and FocalSegNet have more than twice the computational cost of the CNN U-shaped models, with the latter containing the highest number of model parameters. Finally, 3DSwinUnet had the lightest model architecture, and 3DSwinUnetV4 had the highest inference latency on average.

\subsection{Final algorithm ranking}
In our final algorithm ranking for the total anatomical labels, the STUNet, 3DUNet, and AttentionUNet were ranked first, second, and third, respectively. These three models stood out due to their model efficiency and higher segmentation accuracy. Even though 3DUNet ranked second overall, it shared the first place with STUNet for surgical labels and second place with AttentionUNet for BTCV labels. The FocalSegNet, SwinUNETR, 3DSwinUnet, and 3DSwinUnetV4 were ranked next from the fourth to the last position for the combined total anatomical labels. As our ranking balances both accuracy and model complexity, although the baseline 3DSwinUnet had significantly lower segmentation performance (P $<$ 0.05) than the rest, its lowest number of parameters boosted its final ranking. 

\subsection{Ablation studies}

\subsubsection{Impact of attention mechanisms}
Table \ref{table:FourStagePerfRank} provides the segmentation accuracy, model complexity, and ranking of the selected U-shaped models with four resolution stages. While our general observations align with those in Section \ref{SegmentationAccuracy}, the model ranking changed slightly. Among the selected models, 3DUNet ranked first in all of our ranking categories thanks to its computational efficiency and high segmentation accuracy. Following 3DUNet, STUNet came as second in computation complexity, segmentation performance, and final ranking for the surgical labels. Meanwhile, the AttentionUNet ranked as the second best model in our segmentation and final rankings for the BTCV labels and combined labels. Together, FocalSegNet, SwinUNETR, and 3DSwinUnet placed as fourth in the computational complexity category. With the same number of resolution stages, 3DSwinUnet still kept the lowest number of parameters, and FocalSegNet had the lowest inference latency. Finally, FocalSegNet, SwinUNETR, and 3DSwinUnet placed from the fourth to the sixth in segmentation quality and final ranking for all three label groups, respectively.

\subsubsection{Impact of resolution stages}
In Table \ref{table:numStages}, we show the segmentation accuracy results for 3DUNet, STUNet, AttentionUNet, and 3DSwinUnet across different numbers of resolution stages. The findings revealed distinct behaviors among the models when increasing the number of resolution stages. Specifically, increasing the number of resolution stages enhanced the performance of STUNet while it reduced the performance of 3DUNet and 3DSwinUnet. In the case of AttentionUNet, an increase in resolution stages led to a slight enhancement in model performance concerning the NSD score for surgery labels, while its performance was lower in all other scenarios. It is worth noting that despite the minor changes in the model performance due to varying the number of resolution stages, these changes were not statistically significant (p $>$ 0.05).

\begin{table*}[htbp]
  \centering
  \renewcommand{\arraystretch}{2}
  \caption{Model Performance (mean$\pm$std) and ranking across the selected four-stage U-shaped Models. Note that an individual metric's rankings are shown as superscripts beside the corresponding metrics.}
  \label{table:FourStagePerfRank}
  \resizebox{\textwidth}{!}{
    \begin{tabular}{|c|c|c|c|c:c:c|c:c:c|c:c:c|c:c:c|}
      \hline
        \multirow{2}{*}{\textbf{Model Name}} & \multicolumn{3}{c|}{\textbf{Model Complexity $\downarrow$}} & \multicolumn{3}{c|}{\textbf{Dice $\uparrow$}} & \multicolumn{3}{c|}{\textbf{NSD $\uparrow$}} & \multicolumn{3}{c|}{\textbf{Segmentation Rankings $\downarrow$}} & \multicolumn{3}{c|}{\textbf{Final Rankings $\downarrow$}}\\
        \cline{2-16} & Number of Parameter (M) & Inference Latency ($ms \pm ms$) &
        Ranking & BTCV Labels & Surgery Labels & Total & BTCV Labels & Surgery Labels & Total & BTCV & Surgery & Total & BTCV & Surgery & Total \\
      \hline
     $3DUNet$ & $14.59 ^2$ & $5.108 \pm 0.41 ^1$ & 1 & $94.03 \pm 3.02 ^1$ & $97.04 \pm 1.08 ^1$ & $95.41 \pm 1.65 ^1$ & $97.55 \pm 2.58 ^1$ & $98.90 \pm 1.36 ^1$ & $98.17 \pm 1.50 ^1$ & 1 & 1 & 1 & 1 & 1 & 1 \\
    \hline
    $STUNet$ & $14.51 ^1$ & $6.052 \pm 0.65 ^3$ & 2 & $93.75 \pm 3.22 ^3$ & $96.87 \pm 1.14 ^2$ & $95.18 \pm 1.78 ^3$ & $97.41 \pm 2.78 ^3$ & $98.76 \pm 1.37 ^2$ & $98.03 \pm 1.59 ^3$ & 3 & 2 & 3 & 3 & 2 & 3 \\
    \hline
    $AttentionUNet$ & $14.77 ^3$ & $5.970 \pm 0.44 ^2$ & 3 & $93.86 \pm 2.96 ^2$ & $96.82 \pm 1.38 ^3$ & $95.21 \pm 1.67 ^2$ & $97.48 \pm 2.77 ^2$ & $98.72 \pm 1.64 ^3$ & $98.05 \pm 1.70 ^2$ & 2 & 3 & 2 & 2 & 3 & 2 \\
    \hline
    $SwinUNETR$ & $62.19 ^5$ & $18.585 \pm 1.24 ^5$ & 4 & $93.32 \pm 3.21 ^5$ & $96.54 \pm 1.45 ^5$ & $94.79 \pm 1.83 ^5$ & $97.10 \pm 2.86 ^5$ & $98.28 \pm 1.81 ^5$ & $97.64 \pm 1.78 ^5$ & 5 & 5 & 5 & 5 & 5 & 5 \\
    \hline
    $FocalSegNet$ & $69.65 ^6$ & $15.412 \pm 0.84 ^4$ & 4 & $93.47 \pm 2.98 ^4$ & $96.57 \pm 1.39 ^4$ & $94.89 \pm 1.67 ^4$ & $97.20 \pm 2.81 ^4$ & $98.43 \pm 1.58 ^4$ & $97.77 \pm 1.66 ^4$ & 4 & 4 & 4 & 4 & 4 & 4 \\
    \hline
    $3DSwinUnet$ & $30.91 ^4$ & $28.611 \pm 1.32 ^6$ & 4 & $59.71 \pm 6.12 ^6$ & $68.95 \pm 4.56 ^6$ & $63.98 \pm 4.35 ^6$ & $53.96 \pm 6.56 ^6$ & $58.23 \pm 4.59 ^6$ & $55.96 \pm 4.98 ^6$ & 6 & 6 & 6 & 6 & 6 & 6 \\
    \hline
        \end{tabular}%
  }
\end{table*}

\begin{table*}[htbp]
  \centering
  \renewcommand{\arraystretch}{2}
  \caption{Computational complexity and segmentation performance (mean$\pm$std) across 3DUNet, STUNet, AttentionUNet, and 3DSwinUnet models with different numbers of resolution stages.}
  \label{table:numStages}
  \begin{adjustbox}{width=\textwidth}
    \begin{tabular}{|c|c|c|c|c:c:c|c:c:c|}
      \hline
        \multirow{2}{*}{\textbf{Model Name}} & \multirow{2}{*}{\textbf{Num of stages}} & \multicolumn{2}{c|}{\textbf{Model Complexity $\downarrow$}} & \multicolumn{3}{c|}{\textbf{Dice $\uparrow$}} & \multicolumn{3}{c|}{\textbf{NSD $\uparrow$}} \\
        \cline{3-10} & & Number of Parameter (M) & Inference Latency ($ms \pm ms$) & BTCV Labels & Surgery Labels & Total & BTCV Labels & Surgery Labels & Total \\
      \hline
      $3DUNet$ & 4 & 14.59 & $5.108 \pm 0.41$ & $94.03 \pm 3.02$ & $97.04 \pm 1.08$ & $95.41 \pm 1.65$ & $97.55 \pm 2.58$ & $98.90 \pm 1.36$ & $98.17 \pm 1.50$ \\
      \hdashline
      $3DUNet$ & 5 & 30.6 & $6.158 \pm 0.49$ & $93.66 \pm 3.08$ & $96.96 \pm 1.11$ & $95.18 \pm 1.69$ & $97.50 \pm 2.89$ & $98.88 \pm 1.25$ & $98.13 \pm 1.65$  \\
     \hline
      $STUNet$ & 4 & 14.51 & $6.052 \pm 0.65$ & $93.75 \pm 3.22$ & $96.87 \pm 1.14$ & $95.18 \pm 1.78$ & $97.41 \pm 2.78$ & $98.76 \pm 1.37$ & $98.03 \pm 1.59$  \\
      \hdashline
      $STUNet$ & 5 & 30.23 & $7.298 \pm 0.55$ & $94.08 \pm 2.92$ & $97.04 \pm 1.20$ & $95.44 \pm 1.62$ & $97.57 \pm 2.54$ & $98.85 \pm 1.59$ & $98.16 \pm 1.54$  \\
     \hline
      $AttentionUNet$ & 4 & 14.77 & $5.970 \pm 0.44$ & $93.86 \pm 2.96$ & $96.82 \pm 1.38$ & $95.21 \pm 1.67$ & $97.48 \pm 2.77$ & $98.72 \pm 1.64$ & $98.05 \pm 1.70$  \\
      \hdashline
      $AttentionUNet$ & 5 & 30.59 & $7.193 \pm 0.48$ & $93.78 \pm 3.11$ & $96.76 \pm 1.29$ & $95.14 \pm 1.74$ & $97.40 \pm 2.84$ & $98.77 \pm 1.60$ & $98.03 \pm 1.69$   \\
     \hline
      $3DSwinUnet$ & 3 & 7.85 & $22.860 \pm 0.82$ & $60.81 \pm 6.24$ & $70.76 \pm 4.10$ & $65.38 \pm 4.26$ & $54.48 \pm 6.74$ & $60.57 \pm 4.70$ & $57.29 \pm 5.12$ \\
      \hdashline
      $3DSwinUnet$ & 4 & 30.91 & $28.611 \pm 1.32$ & $59.71 \pm 6.12$ & $68.95 \pm 4.56$ & $63.98 \pm 4.35$ & $53.96 \pm 6.56$ & $58.23 \pm 4.59$ & $55.96 \pm 4.98$  \\
     \hline
        \end{tabular}%
  \end{adjustbox}
  
\end{table*}

\subsubsection{3DSwinUnet modifications}
Table \ref{table:SwinUnetVarients} summarizes the computational complexity and segmentation accuracy for the corresponding operations used in each design component of the 3DSwinUnet variants. The findings presented in Table \ref{table:SwinUnetVarients} indicate that replacing linear with convolutional layers can improve model performance. Specifically, replacing the linear layers with residual blocks in skip-connection blocks in 3DSwinUnetV1 led to a significant (p $<$ 0.05) performance improvement compared to 3DSwinUnet. For 3DSwinUnetV2 and 3DSwinUnetV3, replacing linear layers in decoder upsampling blocks with nearest interpolation and transpose convolutional layers, respectively, resulted in a significant (p $<$ 0.05) performance enhancement compared to 3DSwinUnetV1. However, there was only a slight (p $>$ 0.05) difference between the performance of 3DSwinUnetV2 and 3DSwinUnetV3. Finally, using residual layers in encoder downsampling blocks resulted in a slight (p $>$ 0.05) performance improvement in 3DSwinUnetV4 than in 3DSwinUnetV3.

\begin{table*}[htbp]
  \centering
  \renewcommand{\arraystretch}{2}
  \caption{Computational complexity and segmentation performance (mean$\pm$std) across various 3DSwinUnet model variants. Here, SC=skip-connection type, DS=downsampling operation, and US=upsampling operation.}
  \label{table:SwinUnetVarients}
  \resizebox{\textwidth}{!}{
    \begin{tabular}{|c|c|c|c|c:c|c:c:c|c:c:c|}
      \hline
        \multirow{2}{*}{\textbf{Model Name}} & \multicolumn{3}{c|}{\textbf{Details}} & \multicolumn{2}{c|}{\textbf{Computational Cost $\downarrow$}} & \multicolumn{3}{c|}{\textbf{Dice $\uparrow$}} & \multicolumn{3}{c|}{\textbf{NSD $\uparrow$}} \\
        \cline{2-12} & SC & DS & US
        & Number of Parameter (M) & Inference Latency ($ms \pm ms$) &
        BTCV Labels & Surgery Labels & Total & BTCV Labels & Surgery Labels & Total \\
      \hline
      $3DSwinUnet$  & Linear & Linear & Linear & 7.98 & $22.910 \pm 0.93$ & $46.15 \pm 7.02$ & $59.30 \pm 4.60$ & $52.20 \pm 5.06$ & $34.81 \pm 5.64$ & $38.52 \pm 4.15$ & $36.54 \pm 4.24$ \\
      \hline
      $3DSwinUnetV1$ & Residual & Linear & Linear & 24.39 & $27.382 \pm 0.95$ & $86.17 \pm 5.50$ & $92.04 \pm 2.60$ & $88.87 \pm 3.24$ & $89.06 \pm 6.20$ & $91.29 \pm 3.58$ & $90.09 \pm 3.95$ \\
      \hline
      $3DSwinUnetV2$ & Residual & Linear & Interpolation & 23.57 & $27.716 \pm 0.89$ & $91.47 \pm 3.89$ & $95.32 \pm 1.81$ & $93.23 \pm 2.22$ & $95.76 \pm 3.57$ & $96.70 \pm 2.32$ & $96.19 \pm 2.20$ \\
      \hline
      $3DSwinUnetV3$ & Residual & Linear & Transpose Convolution & 24.39 & $27.706 \pm 1.42$ & $91.67 \pm 3.40$ & $95.32 \pm 1.62$ & $93.34 \pm 1.91$ & $95.79 \pm3.36$ & $96.65 \pm 2.40$ & $96.19 \pm 2.11$ \\
      \hline
      $3DSwinUnetV4$ & Residual & Residual & Transpose Convolution & 31.55 & $29.025 \pm 0.81$ & $92.04 \pm 3.58$ & $95.72 \pm 1.58$ & $93.73 \pm 2.00$ & $96.14 \pm 3.47$ & $97.50 \pm 1.96$ & $96.76\pm 2.06$ \\
     \hline
        \end{tabular}%
  }
  
\end{table*}

\section{Discussion}

We compared six state-of-the-art U-shaped techniques and one model variant (3DSwinUnetV4) with a focus on the impacts of different attention mechanisms, including attention gates, self-attention, and focal modulation, as well as comparing CNN-Transformer hybrid and full Transformer models. We didn't observe a significant difference (p$>$0.05) among these models, except for the 3DSwinUnet. However, the attention mechanisms were ranked in the descending order of attention gate, focal modulation, window-based self-attention, and full integration of window-based self-attention. While some previous studies showed the benefit of various attention mechanisms in U-shaped models \cite{tian2019exploration}, in some recent medical image segmentation benchmarking reports \cite{ji2022amos, gut2022benchmarking}, 3D UNets were shown to perform better than CNN-Transformer hybrid counterparts. To further confirm this observation while removing the influence of different resolution stages\cite{kugelman2022comparison}, we conducted an ablation study by fixing the number of resolution stages of all models to four, and the observation remained. This could potentially be explained by the patch-based training and inference, where limited field-of-view and information redundancy may benefit local feature extraction slightly. The segmentation accuracy ranking of the attention mechanisms reflected their ability to encode local features. Furthermore, the observation may also be due to the need for larger training datasets for Transformer models. In clinical deployment, computational efficiency is important, so with similar segmentation accuracy, CNN U-shaped models that contain fewer parameters and faster inference latency can be more favorable. Across all tested models in Table \ref{table:classDice}, we demonstrated that the segmentation quality is in general better for larger anatomies. This is reflected in the significant differences (p$<$0.05) of Dice coefficient and NSD between the BTCV and surgical labels, where the latter contains larger anatomical structures. This observation is consistent with previous studies \cite{ji2022amos}. Furthermore, for these smaller anatomies, the standard deviations of the segmentation accuracy metrics are also greater than those of the larger structures, suggesting lower robustness. 

Deeper layers in convolutional neural networks may help encode more refined features for relevant tasks \cite{tan2019efficientnet}. However, based on Table \ref{table:numStages}, more resolution stages in U-shaped models do not always improve segmentation accuracy. Among the models, STUNet was the only model that slightly benefited from increasing the number of resolution stages potentially because of its extensive incorporation of residual blocks and multi-scale processing in its architecture design. For pure Transformer-based models, increasing the number of resolution stages may reduce performance due to the loss of spatial information in the last layers and cause the attention to collapse \cite{zhou2021deepvit}. To mitigate this, the integration of the attention mechanism with convolutional operations, as proposed in paper \cite{dai2021coatnet} can be beneficial. Finally, as additional resolution stages augment model complexity, the advantage of large models could also depend on the complexity of the task, size of training data, and the initial input image size, and resolution. 

Finally, direct adaptation of the original 2D SwinUnet model \cite{cao2022swin} for 3D segmentation offered sub-optimal performance. Although the shifted window mechanism in the Swin Transformer adds inductive bias to this operation, its self-attention can still interfere with local information, as previously noted in \cite{he2023swinunetr}. The lack of local feature extractors in the 3DSwinUnet, due to the use of Swin Transformer and linear layers, results in suboptimal performance in our analysis. Inspired by the use of residual blocks in the STUNet, we developed four variants of the 3DSwinUNet to fully explore the potential of pure Transformer UNets by modifying the operations of upsampling, skip-connection, and downsampling (Table \ref{table:SwinUnetVarients}). In short, we found greatly enhanced performance when replacing linear operations with residual blocks in skip connections and downsampling operations, as well as employing interpolation or transpose convolution in upsampling operations for the baseline 3DSwinUnet. As residual blocks are known to improve the learning efficiency of hierarchical features and gradient flow during training, they could potentially promote inductive bias compared with simple linear operations. Notably, the performance boost in terms of accuracy and robustness (i.e., lower standard deviations of metrics) was much more evident for their deployment in skip connections than downsampling. While keeping residual blocks for skip-connections and linear operation for downsampling (3DSwinUnetV2 vs. 3DSwinUnetV3), upsampling operations with interpolation and transpose convolution provided similar further accuracy enhancement, with the latter offering slightly better performance robustness. This enhancement was due to the hierarchical representation of convolutional layers \cite{zhou2023nnformer}. 

Our presented study primarily focuses on the impacts of attention mechanisms, number of resolution stages, and network configuration designs for U-shaped deep learning models. As a result, we selected the most popular and representative models for the themes of our investigation, instead of analyzing an exhaustive list of U-shaped models. For the model performance, it is possible that the insights drawn from the experiments may be task- and modality-specific \cite{xiao2023transformers, liu2020survey}, and should be verified further with additional benchmarking datasets.

\section{Conclusion}
In this study, we conducted a comprehensive evaluation of various U-shaped deep learning models in CT-based segmentation for thoracic surgical planning and other abdominal anatomies (BTCV dataset), and showed that the STUNet ranked the best for the designated tasks based on the joint consideration of accuracy and model complexity. In summary, we found that CNN U-shaped models offer excellent values for the demonstrated tasks while attention mechanisms may not necessarily enhance the outcomes, with those better preserving local features gaining a slight edge in patch-based processing. In addition, although augmenting resolution stages does not always result in better accuracy, careful design of operations for different components of the U-shaped models can greatly boost the results. We hope the insights from our experiments will facilitate the deployment and development of deep learning models for the demonstrated application and beyond.

\bibliographystyle{IEEEtran}
\bibliography{refs}

\end{document}